\documentstyle[12pt]{article}
\begin{document}
%%\draft
%%%%%End of Preamble
%%%%Start of Text%%%%%%%%%%%%%%%%%%%%%%%%%%%%%%%%%%%%%%%%%%%%%%%%%%%%%%%

\title{Single chargino production with R-parity lepton number violation
       in photon-photon collisions}
%\footnote{This work was supported by National Natural Science Foundation of
%          China.}}

\vspace{-3mm}
\author{{ Yin Xi$^{a}$, Ma Wen-Gan$^{a}$, Wan Lang-Hui$^{a}$,
          Jiang Yi$^{a}$ and Han Liang$^{a}$}\\
{\small $^{a}$Modern Physics Department, University of Science and}\\
{\small Technology of China, Anhui 230027, China.}}

\date{}
\maketitle \vspace{-4mm}

\begin{abstract}
We examine the process $\gamma \gamma \rightarrow \tilde{\chi}^{+}
\tau^{-}$ at photon-photon collider in the minimal supersymmetric
standard model(MSSM) with R-parity violation, where all the
one-loop diagrams are considered. We mainly discuss the effects of
bilinear breaking terms, and conclude that their contributions may
be important compared with trilinear terms. Our results show that
the events of this process could be detectable at photon-photon
colliders, if the values of the parameters are favorable.
\\
\end{abstract}
%\vspace{20mm}
{\large\bf Key word: R-parity violation, single chargino production, \\
lepton number violation, photon-photon collision}

\vfill \eject

%%\narrowtext

\noindent{\large\bf I. Introduction}

\baselineskip 0.25in
\par
The new physics beyond the standard model (SM) has been
intensively studied in past years\cite{mssm}. The supersymmetric
models (SUSY) are the most attractive ones among general extended
models of the SM. As we know that electroweak gauge invariance
requires the absence of terms in the lagrangian that change either
baryon number or lepton number. Usually these terms may lead to an
unacceptable short proton lifetime. One way to solve the
proton-decay problem is to impose a discrete symmetry conservation
called R-parity ($R_p$) conservation\cite{rparity}. Actually
conservation of R-parity is put into the MSSM in order to ensure
retaining symmetries of the SM. But the most general
supersymmetric (SUSY) extension of the SM should contain such
terms.
\par
If R-parity is conserved, all supersymmetric partner particles
must be produced in pair, thus the lightest of superparticles must
be stable. If R-parity is violated, the feature of SUSY models is
changed a lot. Until now we have being lacking in credible
theoretical argument and experimental tests for $R_p$
conservation, we can say that the $R_p$ violation
(${\rlap/{R}}_p$) would be equally well motivated in the
supersymmetric extension of the SM. Even if we failed in finding a
direct signal of the MSSM in the experiments, it would be also
significant to obtain a signal on $R_p$ violation, which has
recently motivated some investigation\cite{rbreak} \cite{tri}
\cite{rpara} because of experimentally observed discrepancies.
\par
Experimentally searching for the effects of ${\rlap/{R}}_p$
interactions has been done with many efforts in the last few
years. Unfortunately, up to now we have only some upper limits on
${\rlap/{R}}_p$ parameters. It is necessary to continue these
works on finding ${\rlap/{R}}_p$ signal or putting further
stringent constraints on the ${\rlap/{R}}_p$ parameters in future
experiments. The popular way to find a ${\rlap/{R}}_p$ violating
signal is to detect the decay of the lightest supersymmetric
particle(LSP)\cite{lsp}, which is difficult experimentally
especially at photon colliders. The best signature for
${\rlap/{R}}_p$ at future Linear Colliders (LC) is the resonant
Higgs and sneutrino production through the bilinear
${\rlap/{R}}_p$ parameters $\epsilon$ and the trilinear parameters
$\lambda$ or $\lambda'$, respectively. The linear collider
operating in photon collision mode has a distinct advantage over
the situation at the pure $e^{+} e^{-}$ process, where the search
of narrow resonances requires lengthy scans over a large center of
mass energy range of the machine. Similar with the situations at
hadron colliders, the incoming photons (incoming partons at hadron
colliders) with the continuative c.m.s. energy distribution, the
resonance can be probed over a rather wide mass
range\cite{rpara}\cite{wan}. Then a single chargino can be
produced by the Higgs (sneutrino) decays, which can be measured
through the detection of its three-leptons signature. In the
papers of Ref.\cite{rpara}\cite{wan} it is shown that the single
chargino production cross section can be several hundred femto
barn at the LHC.
\par
There are two kinds of ${\rlap/{R}}_p$ parameters, appearing in
the bilinear terms and the trilinear terms respectively. The
bilinear terms account for the mass of neutrinos, and also bring
complexity into the model. Their contribution to ${\rlap/{R}}_p$
process were generally believed to be unimportant, thus simply
neglected in most recent works\cite{tri}. Recently M. Chaichian,
K. Huitu and Z.-H. Yu studied the process $\gamma\gamma
\rightarrow \tilde{\nu} \rightarrow \tilde{\chi}^{\pm}l^{\mp}$
considering only the trilinear ${\rlap/{R}}_p$ terms\cite{yuzh}.
However, in this paper we shall examine the contributions of both
kinds of R-parity violation parameters, and investigate whether
the contribution from bilinear terms can be considerable, even
larger than that from trilinear terms with suitable parameters.
\par
In this paper, we investigate the single chargino production
process via photon-photon collisions at linear colliders (LC)
operating at the energy ranging from 500 GeV to 1 TeV. We discuss
this process in framework of the MSSM with R-parity lepton number
violation. The paper is organized as follows: The general
structure of the model and definitions are presented in Sec II. In
Sec III, there are details of calculation. Numerical results and
discussion are given in Sec IV. In Sec V, there is a short
summary. Appendix A contains some explicit expressions used in our
paper, and finally relevant Feynman rules are collected in
Appendix B.

\begin{flushleft} {\bf II. General Structure of the Model} \end{flushleft}

\par
Supersymmetric Yang-Mills theories contain gauge multiplets $(\lambda^a,
A_\mu^a)$ in the adjoint representation of a gauge group G and matter
multiplets $(A_i, \psi_i)$ in some chosen representations of G. The general
lagrangian of MSSM can be written as
\begin{equation}
\begin{array}{lll}
 {\cal L} &=& -\frac{1}{4} \upsilon_{\mu \nu}^a \upsilon^{a\mu \nu}-
   i\bar{\lambda}^a \bar{\sigma}^\mu D_\mu \lambda^a-\frac{1}{2}
   D^a D^a-F_i^\ast F_i-D_\mu A_i^\ast D^\mu A_i-
   i\bar{\psi}_i\bar{\sigma}^\mu D_\mu \psi_i \\
&& +i\sqrt{2}g(A_i^\ast
   T^a \psi_i \lambda^a-\bar{\lambda}^aT^aA_i\bar{\psi}_i)+
   {\cal L}_{Yukawa}
\end{array}
\end{equation}
where
\begin{equation}
\begin{array}{lll}
&&D_\mu A_i = \partial_\mu A_i+i g A_\mu^aT^aA_i, \\
&&D_\mu \psi_i = \partial_\mu \psi_i+i g A_\mu^aT^a\psi_i, \\
&&D_\mu \lambda^a = \partial_\mu \lambda^a-g f^{abc}A_\mu^b\lambda^c \\
&&\upsilon_{\mu \nu}^a = \partial_\mu A_\nu^a-\partial_\nu A_\mu^a-
        g f^{abc}A_\mu^b A_\nu^c, \\
&&F_i = \left. \frac{\partial {\cal W}}{\partial \Phi_i} \right|_{\theta \theta}, \\
&&D^a = g A_iT_{ij}^aA_j, \\
&&{\cal L}_{Yukawa} = -\frac{1}{2}\left(
    \frac{\partial^2 {\cal W}}{\partial A_i \partial A_j} \psi_i \psi_j +
    h.c. \right)
\end{array}
\end{equation}
\par
$T^a$ is the generator of gauge group G. ${\cal W}$ is
superpotential. Its general form for the R-parity relaxed MSSM can
be written as
\begin{equation}
{\cal W} = {\cal W}_{MSSM}+{\cal W}_{\rlap/{R}}
\end{equation}
where ${\cal W}_{MSSM}$ denotes the R-parity conserved
superpotential
\begin{equation}
{\cal W}_{MSSM} = \mu \epsilon_{ij} H_i^1 H_j^2+\epsilon_{ij} l_I
H_i^1 \tilde{L}_j^I \tilde{R}^I-u_I (H_1^2 C^{\ast JI}
\tilde{Q}_2^J-
     H_2^2 \tilde{Q}_1^J)\tilde{U}^I-d_I (H_1^1 \tilde{Q}_2^I -
     H_2^1 C^{IJ} \tilde{Q}_1^J) \tilde{D}^I
\end{equation}
and ${\cal W}_{\rlap/{R}}$ denotes the terms of R-parity
violation.
\begin{equation}
W_{\rlap/{R}} = \epsilon_{ij} (\lambda_{IJK} \tilde{L}_i^I \tilde{L}_j^J
     \tilde{R}^K+\lambda_{IJK}' \tilde{L}_i^I \tilde{Q}_j^J \tilde{D}^K+
     \epsilon_I H_i^2 \tilde{L}_j^I)+\lambda_{IJK}''\tilde{U}^I
     \tilde{D}^J \tilde{D}^K
\end{equation}
The soft breaking terms can be given by
\begin{equation}
\begin{array}{lll}
{\cal L}_{soft} &=& -m_{H^1}^2 H_i^{1\ast}H_i^1-m_{H^2}^2 H_i^{2\ast} H_i^2-
     m_{L^I}^2 \tilde{L}_i^{I\ast} \tilde{L}_i^I-m_{R^I}^2 \tilde{R}^{I\ast}
     \tilde{R}^I-m_{Q^I}^2 \tilde{Q}_i^{I\ast} \tilde{Q}_i^I \\
&& -m_{D^I}^2
     \tilde{D}^{I\ast} \tilde{D}^I-m_{U^I}^2 \tilde{U}^{I\ast} \tilde{U}^I
     + (m_1 \lambda_B \lambda_B+m_2 \lambda_A^i \lambda_A^i+m_3 \lambda_G^a
     \lambda_G^a+h.c.) \\
&& + \{B \mu \epsilon_{ij} H_i^1 H_j^2+B_I \epsilon_I
     \epsilon_{ij} H_i^2 \tilde{L}_j^I+\epsilon_{ij} l_{sI} H_i^1 \tilde{L}_j^I
     \tilde{R}^I \\
&& +d_{sI}(-H_1^1 \tilde{Q}_2^I+C^{IK} H_2^1 \tilde{Q}_1^K)
     \tilde{D}^I+u_{sI}(-C^{KI\ast} H_1^2 \tilde{Q}_2^I + H_2^2 \tilde{Q}_1^I)
     \tilde{U}^I \\
&& +\epsilon_{ij} \lambda_{IJK}^S \tilde{L}_i^I \tilde{L}_j^J
     \tilde{R}^K+\lambda_{IJK}^{S'}(\tilde{L}_i^I \tilde{Q}_2^J \delta^{JK}-
     \tilde{L}_2^I C^{JK} \tilde{Q}_1^J)\tilde{D}^K+\lambda_{IJK}^{S''}
     \tilde{U}^I \tilde{D}^J \tilde{D}^K \\
&& + h.c.\}
\end{array}
\end{equation}
\par
The bilinear ${\rlap/{R}}_p$ term $\epsilon_{i j} \epsilon_I H_i^2
\tilde{L}_j^I$ will lead to mixture of mass eigenstates in the
$R_p$ violating model\cite{feng}, which would bring complexity
into calculations. The calculation of lagrangian and corresponding
Feynman rules are thus extremely complicated and tedious. The
general process of computation and some useful results are
summarized as following.
\par
With this new model, physical spectrum of particles are quite
different from the $R_p$ conserved theory. The explicit expression
of the mass matrix and eigenstates can be found in
Ref.\cite{feng}. We write down part of the expressions which would
be useful in our following discussion. The fields of two Higgs
doublets $H^1$, $H^2$ and slepton can be written as
\begin{eqnarray}
&&H^1 = \left(
        \begin{array}{c}
        \frac{1}{\sqrt{2}} (\chi_1^0 + i \phi_1^0 + v_1) \\
        H_2^1
        \end{array}
        \right) \\
&&H^2 = \left(
        \begin{array}{c}
        H_1^2 \\
        \frac{1}{\sqrt{2}} (\chi_2^0 + i \phi_2^0 + v_2)
        \end{array}
        \right) \\
&&\tilde{L}^I = \left(
         \begin{array}{c}
         \frac{1}{\sqrt{2}} (\chi_{\nu_I}^0 +
         i \phi_{\nu_I}^0 + v_{\nu_I} )  \\
         \tilde{L}_2^I
         \end{array}
         \right)
\end{eqnarray}
\par
The vacuum expectation values (VEV's) are divided from the scalar fields,
written explicitly as $v_1$, $v_2$ and $v_{\nu_I}$. From the equations above,
we can find the scalar potential including a linear combination as
\begin{equation}
V_{tadpole}=t_1^0 \chi_1^0 + t_2^0 \chi_2^0 + t_{\nu_I}^0 \chi_{\nu_I}^0
\end{equation}
where
\begin{equation}
\begin{array}{lll}
t_1^0 &=& \frac{1}{8} (g^2+g'^2) v_1 (v_1^2-v_2^2+\sum_I
v_{\nu_I}^2) +
          \mid\mu\mid^2 v_1 + m_{H^1}^2 v_1 - B \mu v_2 -
          \sum_{I}^{}\mu\epsilon_I v_I , \\
t_2^0 &=& -\frac{1}{8} (g^2+g'^2) v_2 (v_1^2-v_2^2+\sum_I v_{\nu_I}^2) +
          \mid\mu\mid^2 v_2 + m_{H^2}^2 v_2 - B \mu v_1 +
          \sum_I\epsilon_I^2 v_2 \\
&& + \sum_I B_I\epsilon_I v_I , \\
t_{\nu_I}^0 &=& \frac{1}{8} (g^2+g'^2) v_{\nu_I} (v_1^2-v_2^2+
          \sum_I v_{\nu_I}^2) + m_{L^I}^2 v_{\nu_I} +
          \epsilon_I \sum_J \epsilon_J v_{\nu_J} - \mu\epsilon_I v_1 +
          B_I \epsilon_I v_2 .
\end{array}
\end{equation}
These $t_i^0 (i=1,2,\nu_1,\nu_2,\nu_3)$ correspond to the tree
level tadpoles, and the VEV's of the neutral scalar fields satisfy
the condition $t_i^0=0 (i=1,2,\nu_1,\nu_2,\nu_3)$. Thus the
neutral slepton fields generally obtain none-zero VEV's
$\frac{1}{\sqrt{2}} v_{\nu_I}$, as a result of bilinear terms. The
trilinear parameters alone won't contribute to the VEV's, but to
${\rlap/{R}}_p$ Feynman vertices only.
\par
We use $\chi^0 = (\chi_1^0, \chi_2^0, \chi_{\nu_1}^0,
\chi_{\nu_2}^0, \chi_{\nu_3}^0)$ to denote CP-even neutral scalar
fields and $\phi^0 = (\phi_1^0, \phi_2^0, \phi_{\nu_1}^0,
\phi_{\nu_2}^0, \phi_{\nu_3}^0)$ to represent CP-odd neutral
scalar fields. $\Phi^c = (H_2^{1 *}, H_1^2,\\ \tilde{L}_2^{1 *},
\tilde{L}_2^{2 *}, \tilde{L}_2^{3 *}, \tilde{R}^1, \tilde{R}^2,
\tilde{R}^3)$ denotes the charged scalar fields. The relations
between the current eigenstates and the mass eigenstates are given
as
\begin{eqnarray}
&&H_i^0 = \sum_{j=1}^5 Z_{even}^{i j} \chi_j^0 \\
&&H_{5+i}^0 = \sum_{j=1}^5 Z_{odd}^{i j} \phi_j^0 \\
&&H_i^{+} = \sum_{j=1}^8 Z_c^{i j} \Phi_j^c
\end{eqnarray}
and
\begin{eqnarray}
&&\left( -i\lambda_B, -i\lambda_A^3, \psi_{H_1^1}, \psi_{H_2^2}, \nu_{e L},
         \nu_{\mu L}, \nu_{\tau L} \right)^T = Z_N \tilde{\chi}^0 \\
&&\left( -i\lambda_A^{+}, \psi_{H_1^2}, e_R, \mu_R, \tau_R \right)^T =
         Z_{+} \tilde{\chi}^{+} \\
&&\left( -i\lambda_A^{-}, \psi_{H_2^1}, e_L, \mu_L, \tau_L \right)^T =
         Z_{-} \tilde{\chi}^{-}
\end{eqnarray}
\par
We may formulate all the neutral fermions into four-component
spinors as
\begin{equation}
\kappa_i^0 = \left(
             \begin{array}{ll}
             \tilde{\chi}_i^0 \\
             \bar{\tilde{\chi}}_i^0
             \end{array}
             \right)
\end{equation}
and the charged fermions as
\begin{equation}
\kappa_i^{+} = \left(
               \begin{array}{ll}
               \tilde{\chi}_i^{+} \\
               \bar{\tilde{\chi}}_i^{-}
               \end{array}
               \right),
\kappa_i^{-} = \left( \kappa_i^{+} \right)^c =
               \left(
               \begin{array}{ll}
               \tilde{\chi}_i^{-} \\
               \bar{\tilde{\chi}}_i^{+}
               \end{array}
               \right)
\end{equation}
\par
For convenience we will call all of these scalar bosons ($H^1$, $H^2$ and
$\tilde{L}^I$) as Higgs ($H^0$ for CP-even Higgs, $A^0$ for CP-odd Higgs, and
$H^{\pm}$ for charged Higgs), all charged fermions (chargino and
lepton) as chargino ($\tilde{\chi}^{+}$), and all neutral fermions (neutralino and
neutrino) as neutralino ($\tilde{\chi}^0$).
\par
We adopt the 't Hooft-Feynman gauge, in which the Goldstone fields
appear explicitly in the calculations, but ghost vertices are
relatively simple. The lagrangian in terms of physical eigenstates
is very complicated and tedious. For the convenience of
calculation we collect the relevant Feynman rules in Appendix B.

\begin{flushleft} {\bf III. Calculations} \end{flushleft}
\par
In this paper we denote the process
\begin{equation}
\gamma (k_1) \gamma (k_2) \longrightarrow \tilde{\chi}^{+}(p_1) \tau^{-}(p_2)
\end{equation}
According to the MSSM of R-parity violation discussed above, there
is no Feynman vertices of photon-chargino-lepton, and thus the
lowest order diagrams are of one-loop. In this case it's not
necessary to consider the renormalization at one-loop level and
the ultraviolet divergence should be canceled automatically if all
the one-loop diagrams at the $O(m_{\tilde{\chi}_1^{+}}
m_{\tau}/m_W^2)$ order in the model are included. The one-loop
diagrams at the lowest order are plotted in Fig.1, where the
conventions claimed in Sec II is used, such as $\kappa_i^{+}$
represent both chargino and lepton and so on. The diagrams
exchanging the two external photon-photon lines are not shown.
Fig.1(a) contains the self-energy diagrams, Fig.1(b) the vertex
correction diagrams, Fig.1(c) the s-channel diagrams, Fig.1(d) the
box-diagrams, and Fig.1(e) the quartic interaction diagrams.
\par
In the calculation for the s-channel diagrams (Fig.1 c), we take
into account the width effects of the $H^0$, $A^0$ and sneutrino
propagators. The full calculation of mass width is complicated in
the $R_{p}$-violating MSSM model. However, rough results can be
obtained by some analysis. Since we set the mass of sneutrino
around 400 GeV, the $t\bar{t}$ decay needs to be considered. On
the other hand, for $H^0$ and $A^0$, $m_{H^0}, m_{A^0}<2 m_t$, the
decay channel to $t\bar{t}$ is forbidden. Thus we consider the
decay widths for sneutrinos are larger than those of Higgs, due to
the large mass of $m_t$. For an approximate calculation, we choose
$\frac{\Gamma (sneutrino)} {M(sneutrino)} = 0.15$, while
$\frac{\Gamma (H^0)}{m_{H^0}} = \frac{ \Gamma (A^0)}{m_{A^0}} =
0.10$ properly.
\par
We denote $\theta$ as the scattering angle between one of the photons and the
final chargino. Then we express all the four-momenta of the initial and final
particles in the center-of-mass (CMS) by means of the total energy
$\sqrt{\hat{s}}$ and the scattering angle $\theta$. The four-momenta of
chargino and tau are $p_1$ and $p_2$ respectively and are read
\begin{eqnarray}
&&p_1 = \left( E_{\tilde{\chi}}, \sqrt{E_{\tilde{\chi}}^2-m_{\tilde{\chi}}^2}
      \sin\theta, 0, \sqrt{E_{\tilde{\chi}}^2-m_{\tilde{\chi}}^2}\cos\theta
      \right) \\
&&p_2 = \left( E_{\tau}, -\sqrt{E_{\tau}^2-m_{\tau}^2}\sin\theta, 0,
      \sqrt{E_{\tau}^2-m_{\tau}^2}\cos\theta \right)
\end{eqnarray}
where
\begin{eqnarray}
&&E_{\tilde{\chi}} = \frac{1}{2}(\sqrt{\hat{s}}+
             (m_{\tilde{\chi}}^2-m_{\tau}^2)/\sqrt{\hat{s}}) \\
&&E_{\tau} = \frac{1}{2}(\sqrt{\hat{s}}-
             (m_{\tilde{\chi}}^2-m_{\tau}^2)/\sqrt{\hat{s}})
\end{eqnarray}
$p_3$ and $p_4$ are the four-momenta of the initial photons and are expressed
as
\begin{eqnarray}
&&p_3 = (\frac{1}{2}\sqrt{\hat{s}}, 0, 0, \frac{1}{2}\sqrt{\hat{s}}) \\
&&p_4 = (\frac{1}{2}\sqrt{\hat{s}}, 0, 0, -\frac{1}{2}\sqrt{\hat{s}})
\end{eqnarray}
\par
The corresponding matrix element for all the diagrams in Fig.1 is written
in the form
\begin{equation}
\begin{array}{lll}
M &=& M^s + M^{tr} + M^b + M^q
  = M^{s,\hat{t}} + M^{s,\hat{u}} + M^{tr,\hat{t}} + M^{tr,\hat{u}} + M^q \\
  &=& \epsilon^{\mu}(p_1) \epsilon^{\nu}(p_2) \bar{u}(k_1) \{
   f_1 g_{\mu \nu}+
   f_2 \gamma_5 g_{\mu \nu}+
   f_3 \gamma_{\mu} k_{1,\nu}+
   f_4 g_{\mu \nu} \rlap/p_2+
   f_5 k_{1,\nu} \gamma_5  \gamma_{\mu}+ \\
&& f_6 g_{\mu \nu} \gamma_5  \rlap/p_2+
   f_7 \gamma_{\mu}  \gamma_{\nu}+
   f_8 \gamma_5  \gamma_{\mu} \gamma_{\nu}+
   f_9 \gamma_{\mu}  \gamma_{\nu} \rlap/p_2+
   f_{10} \gamma_5 \gamma_{\mu}  \gamma_{\nu}  \rlap/p_2+
   f_{11} k_{1,\mu} k_{1,\nu}+ \\
&& f_{12} \gamma_5 k_{1,\mu} k_{1,\nu}+
   f_{13} \gamma_{\nu} k_{1,\mu}+
   f_{14} g_{\mu \nu} \rlap/p_1+
   f_{15} k_{1,\mu} \gamma_5  \gamma_{\nu}+
   f_{16} g_{\mu \nu} \gamma_5 \rlap/p_1+
   f_{17} k_{1,\nu} \gamma_{\mu}  \rlap/p_1+ \\
&& f_{18} k_{1,\mu} \gamma_{\nu}  \rlap/p_2+
   f_{19} g_{\mu \nu} \rlap/p_1  \rlap/p_2+
   f_{20} k_{1,\nu} \gamma_5  \gamma_{\mu}  \rlap/p_1+
   f_{21} k_{1,\mu} \gamma_5 \gamma_{\nu} \rlap/p_2+
   f_{22} g_{\mu \nu} \gamma_5 \rlap/p_1  \rlap/p_2+ \\
&& f_{23} \gamma_{\mu}  \gamma_{\nu}  \rlap/p_1+
   f_{24} \gamma_5  \gamma_{\mu}  \gamma_{\nu}  \rlap/p_1+
   f_{25} \gamma_{\mu} \gamma_{\nu} \rlap/p_1  \rlap/p_2+
   f_{26} \gamma_5  \gamma_{\mu}  \gamma_{\nu}  \rlap/p_1 \rlap/p_2+
   f_{27} k_{1,\mu} k_{1,\nu} \rlap/p_1+ \\
&& f_{28} k_{1,\mu} k_{1,\nu} \gamma_5  \rlap/p_1+
   f_{29} k_{1,\mu} \gamma_{\nu} \rlap/p_1+
   f_{30} k_{1,\mu} \gamma_5  \gamma_{\nu}  \rlap/p_1+
   f_{31} k_{1,\mu} \gamma_{\nu} \rlap/p_1  \rlap/p_2+
   f_{32} k_{1,\mu} \gamma_5  \gamma_{\nu}  \rlap/p_1  \rlap/p_2+ \\
&& f_{33} \gamma_{\mu} p_{1,\nu}+
   f_{34} \gamma_5  \gamma_{\mu} p_{1,\nu}+
   f_{35} k_{1,\mu} p_{1,\nu}+
   f_{36} \gamma_5 k_{1,\mu} p_{1,\nu}+
   f_{37} \gamma_{\mu} \rlap/p_1 p_{1,\nu}+ \\
&& f_{38} \gamma_{\mu} \rlap/p_2 p_{1,\nu}+
   f_{39} \gamma_5 \gamma_{\mu}  \rlap/p_1 p_{1,\nu}+
   f_{40} \gamma_5 \gamma_{\mu}  \rlap/p_2 p_{1,\nu}+
   f_{41} k_{1,\mu} k_{1,\nu} \rlap/p_2+
   f_{42} k_{1,\mu} k_{1,\nu} \gamma_5  \rlap/p_2+ \\
&& f_{43} k_{1,\nu} \gamma_{\mu}  \rlap/p_2+
   f_{44} k_{1,\nu} \gamma_5 \gamma_{\mu}  \rlap/p_2+
   f_{45} k_{1,\nu} \gamma_{\mu}  \rlap/p_1  \rlap/p_2+
   f_{46} k_{1,\nu} \gamma_5  \gamma_{\mu} \rlap/p_1  \rlap/p_2+
   f_{47} k_{1,\mu} p_{1,\nu} \rlap/p_2+ \\
&& f_{48} k_{1,\mu} \gamma_5  \rlap/p_2 p_{1,\nu}+
   f_{49} \gamma_{\mu}  \rlap/p_1  \rlap/p_2 p_{1,\nu}+
   f_{50} \gamma_5  \gamma_{\mu}  \rlap/p_1  \rlap/p_2 p_{1,\nu}+
   f_{51} g_{\mu \nu} \gamma_5  \rlap/p_1+
   f_{52} k_{1,\mu} p_{1,\nu} \rlap/p_1+ \\
&& f_{53} k_{1,\mu} \gamma_5  \rlap/p_1 p_{1,\nu}+
   f_{54} g_{\mu \nu} \rlap/p_1 \rlap/p_2+
   f_{55} g_{\mu \nu} \gamma_5  \rlap/p_1  \rlap/p_2+
   f_{56} \epsilon_{\alpha \beta \mu \nu} p_{1}^{\alpha} p_{2}^{\beta}+
   f_{57} \gamma_5 \epsilon_{\alpha \beta \mu \nu} p_{1}^{\alpha} p_{2}^{\beta}+ \\
&& f_{58} \epsilon_{\alpha \beta \mu \nu} \gamma^{\alpha}
p_{1}^{\beta}+
   f_{59} \epsilon_{\alpha \beta \mu \nu} \gamma^{\alpha} p_{2}^{\beta}+
   f_{60} \epsilon_{\alpha \beta \mu \nu} p_{1}^{\alpha} p_{2}^{\beta} \rlap/p_1+
   f_{61} \epsilon_{\alpha \beta \mu \nu} p_{1}^{\alpha} p_{2}^{\beta} \rlap/p_2+ \\
&& f_{62} \gamma_5  \epsilon_{\alpha \beta \mu \nu}
\gamma^{\alpha} p_{1}^{\beta}+
   f_{63} \gamma_5  \epsilon_{\alpha \beta \mu \nu} \gamma^{\alpha} p_{2}^{\beta}+
   f_{64} \epsilon_{\alpha \beta \mu \nu} p_{1}^{\alpha} p_{2}^{\beta} \gamma_5 \rlap/p_1+
   f_{65} \epsilon_{\alpha \beta \mu \nu} p_{1}^{\alpha} p_{2}^{\beta} \gamma_5 \rlap/p_2+ \\
&& f_{66} \epsilon_{\alpha \beta \gamma \mu} \gamma^{\alpha}
p_{1}^{\beta} p_{2}^{\gamma} p_{1,\nu}+
   f_{67} \gamma_5 \epsilon_{\alpha \beta \gamma \mu} \gamma^{\alpha}
            p_{1}^{\beta} p_{2}^{\gamma} p_{1,\nu}+
   f_{68} \gamma_{\nu} k_{2,\mu}+
   f_{69} k_{2,\mu} \gamma_5 \gamma_{\nu}+
   f_{70} \gamma_{\mu} k_{2,\nu}+ \\
&& f_{71} k_{2,\nu} \gamma_5 \gamma_{\mu}
     \} v(k_2)
\end{array}
\end{equation}
\par
The variables $\hat{s}$, $\hat{t}$ and $\hat{u}$ are usual Mandelstam variables
in the center of mass system of photon-photon, defined as
\begin{eqnarray}
&&\hat{s} = (p_1+p_2)^2 = (p_3+p_4)^2 \\
&&\hat{t} = (p_1-p_3)^2 = (p_2-p_4)^2 \\
&&\hat{u} = (p_1-p_4)^2 = (p_2-p_3)^2
\end{eqnarray}
\par
Explicit expressions of the factors $f_i$ appearing in Eq.(27) are
very complicated and tedious, thus we do not list them in our
paper. The total cross section for the subprocess $\gamma \gamma
\rightarrow \tilde{\chi}^{+} \tau^{-}$ can be written in the form
as
\begin{equation}
\hat{\sigma}(\hat{s}) = \frac{1}{16\pi\hat{s}^2}\int_{\hat{t}^{-}}^
             {\hat{t}^{+}}d\hat{t}\mid\bar{M}\mid^2
\end{equation}
where $\mid\bar{M}\mid^2$ is the initial spin-averaged matrix element squared
and $\hat{t}^{\pm} = \frac{1}{2}(m_{\tilde{\chi}}^2+m_{\tau}^2-\hat{s})\pm
\sqrt{E_{\tilde{\chi}}^2-m_{\tilde{\chi}}^2}\sqrt{\hat{s}}$.
\par
With the integrated photon luminosity in the $e^{+}e^{-}$
collisions, the total cross section of the process
$e^{+}e^{-}\rightarrow\gamma\gamma\rightarrow
\tilde{\chi}^{+}\tau^{-}$ can be written as
\begin{equation}
\sigma(s) = \int_{(m_{\tilde{\chi}}+m{\tau})/\sqrt{\hat{s}}}^{x_{max}} dz
     \frac{d{\cal L}_{\gamma \gamma}}{dz} \hat{\sigma}
     (\gamma\gamma\rightarrow\tilde{\chi}^{+}\tau^{-} at \hat{s}=z^2 s)
\end{equation}

where $\sqrt{s}$ and $\sqrt{\hat{s}}$ are the total energy of the
center-of-mass of $e^{+}e^{-}$ and $\gamma \gamma$ respectively.
$\frac{d{\cal L}_{\gamma\gamma}} {dz}$ is defined as
\begin{equation}
\frac{d{\cal L}_{\gamma\gamma}}{dz} = 2z \int_{z^2/x_{max}}^{x_{max}} \frac{dx}
       {x} F_{\gamma/e}(x) F_{\gamma/e}(z^2/x)
\end{equation}
\par
For the initial unpolarized electrons and laser photon beams, the energy
spectrum of the back-scattered is given by\cite{energyspectrum}
\begin{equation}
F_{\gamma/e}=\frac{1}{D(\xi)} \left[1-x+\frac{1}{1-x}-\frac{4x}{\xi (1-x)}+
       \frac{4x^2}{\xi^2 (1-x)^2} \right]
\end{equation}
where
\begin{eqnarray}
D(\xi)=(1-\frac{4}{\xi}-\frac{8}{\xi^2}) \ln (1+\xi)+\frac{1}{2}+\frac{8}{\xi}
       -\frac{1}{2(1+\xi)^2} \\
\xi=4 E_0 \omega_0 / m_e^2
\end{eqnarray}
where $m_e$ and $E_0$ are the mass and energy of the electron
respectively, $x$ represents the fraction of the energy of the
incident electron carried by the back-scattered photon. In our
evaluation, we choose $\omega_0$ such that it maximizes the
backscattered photon energy without spoiling the luminosity by
$e^{+}e^{-}$ pair production. Then we get $\xi=2(1+\sqrt{2})\simeq
4.8, x_{max}\simeq 0.83$ and $D(\xi)\simeq 1.8$, a usual method
used in Ref.\cite{eerr}.

\vskip 5mm
\noindent{\large\bf IV. Numerical Calculations and Discussions}
\par
In the numerical computation we use input parameters as $m_b=4.5
GeV, m_c=1.35 GeV, m_t=175 GeV, M_W=80.2226 GeV,
G_F=1.166392\times 10^{-5} (GeV)^{-2}$ and
$\alpha=1/137.036$\cite{inputpara}.
\par
Although there are some constraints on the supersymmetric
parameters in some theory, such as the minimal supergravity
(mSUGRA), in the following analysis we do not put any extra
limitations on them for the general discussion. In the numerical
calculation, we set the following parameters arbitrarily, in case
that no special declaration has been presented on them.

\begin{equation}
\begin{array}{l}
      \tan\beta=5,~~~m_{L^I}=400~GeV \\
      m_{Q^{1,2}}=m_{D^{1,2}}=m_{U^{1,2}}=600~GeV \\
      m_{Q^3}=400~GeV,~~m_{D^3}=500~GeV,~~m_{U^3}=300~GeV
\end{array}
\end{equation}
where $m_{L^I},m_{Q^I},m_{D^I},m_{U^I}$ are the soft-breaking masses appeared
in Eq.(6). We choose $m_2$ in the range from 45 GeV to 150 GeV, and $\mu$
from 100 GeV to 370 GeV. If we omit $R_p$ breaking, the chargino mass matrix
can be written as

\begin{equation}
M_{\tilde{\chi}^{+}}=\left(
\begin{array}{cc}
2 m_2                     &  \frac{1}{\sqrt{2}} g v_1 \\
\frac{1}{\sqrt{2}} g v_2  &  -\mu
\end{array}
\right)
\end{equation}
In the case of large $\mu$, mass of $\tilde{\chi}_1^{+}$ is close
to $2 m_2$. The ${\rlap/{R}}_p$ parameters are taken as follows

\begin{equation}
\begin{array}{l}
 \epsilon_1=\epsilon_2=0,~~\epsilon_3=10~GeV \\
 v_{\nu_1}=v_{\nu_2}=0,~~v_{\nu_3}=-4.7~GeV
\end{array}
\end{equation}
and

\begin{equation}
\begin{array}{l}
 \lambda_{131}=0.05,~~\lambda_{232}=0.09 \\
 \lambda'_{ijk}=0.1,~~\lambda''_{ijk}=0
\end{array}
\end{equation}
\par
The first two indices for parameters $\lambda_{ijk}$ are
antisymmetric, or written explicitly,
$\lambda_{ijk}=-\lambda_{jik}$. For the rest of parameters which
are not given in Eq.(38), we set $\lambda_{ijk}=0$ assuming the
${\rlap/{R}}_p$ for the first two generations can be neglected.
Since we mainly investigate the effects of bilinear breaking
parameters $\epsilon$, trilinear parameters are generally assumed
to be vanished without emphasis.
\par
In Fig.2 we depict the dependences of $\gamma\gamma \rightarrow
\tilde{\chi}^{+} \tau^{-}$ subprocess cross section on the c.m.
energy $\sqrt{\hat{s}}$, where the contributions from all diagrams
with effects of only bilinear terms are included. The masses of
the final chargino are set as 90 GeV, 110 GeV and 170 GeV
respectively. Generally two peaks at $\sqrt{\hat{s}}\sim
m_{H^0}\simeq 290~GeV$ and $\sqrt{\hat{s}}\sim
m_{\tilde{\nu}}\simeq 390~GeV$ can be seen, as the result of
resonance effects through decays of $H^0$, $A^0$ and
$\tilde{\nu}$. When the mass of $\tilde{\chi}_1$ decreases to 90
GeV, which is close to the mass of $h^0$, the cross section is
enhanced considerably by resonance effect. The small peak of the
solid line, where $\sqrt{\hat{s}}\sim 2 m_W, 2 m_{h^0} \simeq
174~GeV$, comes from the contribution of the resonant effects of
box and quartic vertex diagrams including $W^{+}$ and $h^0$
particles. For comparison we also calculated the contributions of
the diagrams with the couplings induced from only the trilinear
breaking terms. In this case neutral Higgs decay to
$\tilde{\chi}_1^{+}\tau^{-}$ is forbidden, thus we take into
account only $\tilde{\nu}$ decay channel. Their contributions can
be typically 1-2 orders lower than those of bilinear terms, which
is not plotted explicitly.
\par
The cross sections of the
$e^{+}e^{-}\rightarrow\gamma\gamma\rightarrow
\tilde{\chi}_1^{+}\tau^{-}$ process and the $e^{+}e^{-}\rightarrow
\tilde{\chi}_1^{+}\tau^{-}$ tree level process are plotted as
functions of $\sqrt{s}$ in Fig.3 neglecting the contributions from
the trilinear terms. Fig.3 shows that when $m_{
\tilde{\chi}_1}=90GeV$, the $\gamma\gamma$ one-loop cross section
has its maximum at 0.39 fb. The cross section of the $e^{+}e^{-}$
tree level process has been calculated in Ref.\cite{feng}.
According to our results in Fig.3, $\sigma_{Born}(e^{+}e^{-}
\rightarrow\tilde{\chi}_1^{+}\tau^{-})$ ranges from 0.1 fb to 0.5
fb when mass of $\tilde{\chi}_1^{+}$ varies from 190 GeV to 90
GeV. We have $\sigma(e^{+}e^{-}\rightarrow\gamma\gamma\rightarrow
\tilde{\chi}_1^{+}\tau^{-})
/\sigma(e^{+}e^{-}\rightarrow\tilde{\chi}_1^{+} \tau^{-})\simeq
10^{-1} - 1$. It shows that contributions of the one-loop level
$\gamma\gamma$ process can be comparable with the tree level
$e^{+}e^{-}$ process, thus a possible way to detect single
chargino production.
\par
In Fig.4 we depict the cross section of the $e^{+}e^{-}$ process
versus the final chargino mass $m_{\tilde{\chi}_1}$, where
$\mu=370~GeV$. We choose the c.m. collision energy $\sqrt{s}$ at
500 GeV and 1000 GeV, and $m_{\tilde{\chi}_1}$ ranges from 90 GeV
to 200 GeV. The cross section decreases as $m_{\tilde{\chi}_1}$
rises, ranging from 0.39 fb to 0.01 fb.
\par
The dependence of the cross sections on parameter $\mu$ is plotted
in Fig.5, where $m_2=150~GeV$. We choose $\sqrt{s}$ at 500 GeV and
1000 GeV respectively. The variety of $\mu$ has effects on
$m_{H^0}$ and $m_{\tilde{\chi}_1}$, thus on the contribution of
resonance effects. While $\mu$ ranges from 100 GeV to 300 GeV, the
cross section starts at 0.61 fb ($\sqrt{s}=500~GeV$), drops
rapidly till $\mu=150~GeV$, then decreases slower, ended at 0.008
fb.
\par

\vskip 10mm
\noindent
{\Large{\bf V.Summary}}
\vskip 5mm
\par
  In this work we have studied the ${\rlap/{R}}_p$ process $e^{+}e^{-}
\rightarrow \gamma \gamma \rightarrow \tilde{\chi}_1^{+} \tau^{-}$
in the framework of R-parity relaxed MSSM, concerning mainly
bilinear parameters. We investigated the dependence of the cross
section on c.m. energy $\sqrt{s}$, mass of $\tilde{\chi}_1^{+}$
and $\mu$. Contributions of bilinear breaking terms and trilinear
terms with typical parameters are compared. Our results show that
it is contrary to the general expectation, bilinear terms account
for most of the contribution in our calculation, which bring more
diagrams into consideration, including neutral Higgs s-channel
diagrams. For LC operating at c.m.s. energy of 500 GeV with
$50~fb^{-1}$ integrated luminosity, one can expect a typical rate
of 11 raw events when $M_{\tilde{\chi}_1}=100 GeV$. That means the
$\tilde{\chi}_1^{+}\tau^{-}$ production may be detectable at
future LC.

\vskip 5mm \noindent{\large\bf Acknowledgement:} This work was
supported in part by the National Natural Science Foundation of
China(project numbers: 19675033,10005009), the Education Ministry
of China and the State Commission of Science and Technology of
China.

\vskip 5mm
\noindent{\large\bf Appendix A: Definitions in the Calculation}

\begin{eqnarray*}
A_{ve}^i&=&v_1 Z_{even}^{i,1}-v_2 Z_{even}^{i,2}+v_{\nu_I} Z_{even}^{i,2+I} \\
A_{ec}^{ij}&=&Z_{even}^{i,1}Z_c^{j,1}+Z_{even}^{i,2}Z_c^{j,2}+
            Z_{even}^{i,2+I}Z_c^{j,2+I} \\
A_{oc}^{ij}&=&Z_{odd}^{i,1}Z_c^{j,1}-Z_{odd}^{i,2}Z_c^{j,2}+
            Z_{odd}^{i,2+I}Z_c^{j,2+I} \\
A_{cc}^{ij}&=&Z_c^{i,1}Z_c^{j,1}-Z_c^{i,2}Z_c^{j,2}+Z_c^{i,2+I}Z_c^{j,2+I} \\
A_{ecc}^{ijk}&=&\frac{1}{4} \{-(g^2-g'^2)A_{cc}^{jk}A_{ve}^i +
        g^2 [v_1(A_{ec}^{ik}Z_c^{j,1}+A_{ec}^{ij}Z_c^{k,1}) +
        v_2(A_{ec}^{ik}Z_c^{j,2}+A_{ec}^{ij}Z_c^{k,2}) \\
&&      + v_{\nu_I}(A_{ec}^{ik}Z_c^{j,2+I}+A_{ec}^{ij}Z_c^{k,2+I})] -
        2g'^2 A_{ve}^i Z_c^{j,5+I}Z_c^{k,5+I} \\
&&      - 2 l_I^2 [ (Z_c^{j,2+I}Z_c^{k,1}
        + Z_c^{j,1}Z_c^{k,2+I}) (v_{\nu_I}Z_{even}^{i,1}
        +v_1 Z_{even}^{i,2+I}) \\
&&      - 2 v_1(Z_c^{j,2+I}Z_c^{k,2+I}+Z_c^{j,5+I}Z_c^{k,5+I})Z_{even}^{i,1}
        -2 v_{\nu_I}Z_c^{j,1}Z_c^{k,1}Z_{even}^{i,2+I} ] \\
&&      -2 \sqrt{2} \mu l_{sI} [
        (Z_c^{k,5+I}Z_c^{j,2+I}+Z_c^{k,2+I}Z_c^{j,5+I})Z_{even}^{i,1}
        -(Z_c^{j,5+I}Z_c^{j,1}+Z_c^{k,1}Z_c^{j,5+I})Z_{even}^{i,2+I} ] \\
&&      + 2 \sqrt{2} \mu l_{I} [
        (Z_c^{k,5+I}Z_c^{j,2+I}+Z_c^{k,2+I}Z_c^{j,5+I})Z_{even}^{i,2}
        + (Z_c^{k,5+I}Z_c^{j,2}+Z_c^{k,2}Z_c^{j,5+I})Z_{even}^{i,2+I}] \\
&&      + 2 \sqrt{2} l_I \epsilon_I [
        (Z_c^{k,5+I}Z_c^{j,1}+Z_c^{k,1}Z_c^{j,5+I})Z_{even}^{i,2}
        +(Z_c^{k,5+I}Z_c^{j,2}+Z_c^{k,2}Z_c^{j,5+I})Z_{even}^{i,1}] \\
&&      + 2 l_I l_J Z_c^{j,5+J}Z_c^{k,5+I}
        (v_{\nu_J}Z_{even}^{i,2+I}+v_{\nu_I}Z_{even}^{i,2+J}) \} \\
A_{occ}^{ijk}&=&\frac{1}{4} \{g^2 [v_1(A_{oc}^{ik}Z_c^{j,1} -
        A_{oc}^{ij}Z_c^{k,1}) +
        v_2(A_{oc}^{ik}Z_c^{j,2}-A_{oc}^{ij}Z_c^{k,2}) \\
&&      + v_{\nu_I}(A_{oc}^{ik}Z_c^{j,2+I}-A_{oc}^{ij}Z_c^{k,2+I})] \\
&&      + 2 l_I^2 (Z_c^{j,2+I}Z_c^{k,1}-Z_c^{j,1}Z_c^{k,2+I})
        (-v_{\nu_I}Z_{odd}^{i,1}
        + v_1 Z_{odd}^{i,2+I}) \\
&&      - 2 \sqrt{2} \mu l_{sI} [
        (Z_c^{k,5+I}Z_c^{j,2+I}-Z_c^{k,2+I}Z_c^{j,5+I})Z_{odd}^{i,1}
        -(Z_c^{k,5+I}Z_c^{j,1}-Z_c^{k,1}Z_c^{j,5+I})Z_{odd}^{i,2+I}] \\
&&      - 2 \sqrt{2} \mu l_{I} [
        (Z_c^{k,5+I}Z_c^{j,2+I}-Z_c^{k,2+I}Z_c^{j,5+I})Z_{odd}^{i,2}
        -(Z_c^{k,5+I}Z_c^{j,2}-Z_c^{k,2}Z_c^{j,5+I})Z_{odd}^{i,2+I}] \\
&&      - 2 \sqrt{2} l_I \epsilon_I [
        (Z_c^{k,5+I}Z_c^{j,1}-Z_c^{k,1}Z_c^{j,5+I})Z_{odd}^{i,2}
        -(Z_c^{k,5+I}Z_c^{j,2}-Z_c^{k,2}Z_c^{j,5+I})Z_{odd}^{i,1}] \\
&&      + 2 l_I l_J Z_c^{j,5+J}Z_c^{k,5+I}
        (v_{\nu_J}Z_{odd}^{i,2+I}-v_{\nu_I}Z_{odd}^{i,2+J}) \}
\end{eqnarray*}

\vskip 5mm
\noindent{\large\bf Appendix B: Relevant Feynman Vertices}

We write down all the relevant ${\rlap/{R}}_p$ Feynman vertices as follows

\begin{eqnarray*}
H_i^0-W_{\mu}^{-}-W_{\nu}^{+}&:&
-\frac{i}{2}g^2 A_{ve}^i g_{\mu \nu} \\
H_1^{+}-W_{\mu}^{-}-A_{\nu}&:&
\frac{i}{2} \sqrt{g^2+g'^2} m_W \sin 2\theta_W g_{\mu \nu} \\
H_1^{+}-W_{\mu}^{+}-Z_{\nu}^0&:&
\frac{i}{2} \sqrt{g^2+g'^2} m_W (\cos 2\theta_W-1) g_{\mu \nu} \\
H_i^0(p_1)-\bar{H}_j^{+}(p_2)-W_{\mu}^{+}&:&
-\frac{i}{2}g A_{ec}^{i j} (p_1+p_2)^{\mu} \\
H_{5+i}^0(p_1)-\bar{H}_j^{+}(p_2)-W_{\mu}^{+}&:&
-\frac{1}{2}g A_{oc}^{i j} (p_1+p_2)^{\mu} \\
\bar{H}_i^{+}(p_1)-H_j^{+}(p_2)-A_{\mu}&:&
\frac{i}{2} \sqrt{g^2+g'^2} \sin 2\theta_W \delta_{i j} (p_1+p_2)^{\mu} \\
\bar{H}_i^{+}(p_1)-H_j^{+}(p_2)-Z_{\mu}^0&:&
\frac{i}{2} \sqrt{g^2+g'^2} (\cos 2\theta_W \delta_{i j} -
            Z_c^{i,5+I} Z_c^{j,5+I}) (p_1+p_2)^{\mu} \\
H_i^0-\bar{H}_j^{+}-A_{\mu}-W_{\nu}^{+}&:&
\frac{i}{2}g^2 \sin\theta_W A_{ec}^{i j} g_{\mu \nu} \\
H_{5+i}^0-\bar{H}_j^{+}-A_{\mu}-W_{\nu}^{+}&:&
\frac{1}{2}g^2 \sin\theta_W A_{oc}^{i j} g_{\mu \nu} \\
\bar{H}_i^{+}-H_j^{+}-A_{\mu}-A_{\nu}&:&
-\frac{i}{2} (g^2+g'^2) \sin^2 2\theta_W \delta_{i j} g_{\mu \nu} \\
\bar{H}_i^{+}-H_j^{+}-A_{\mu}-Z_{\nu}^0&:&
-\frac{i}{2} (g^2+g'^2) [\cos 2\theta_W \sin 2\theta_W \delta_{i j} \\
&& - Z_c^{i,5+I} Z_c^{j,5+I} (\cos 2\theta_W \sin 2\theta_W +
    4 \sin^3 \theta_W \cos \theta_W)] g_{\mu \nu} \\
W_{\mu}^{+}-\kappa_j^0-\bar{\kappa}_i^{+}&:&
ig\gamma^{\mu} [(Z_{+}^{*i,1} Z_N^{j,2}-\frac{1}{\sqrt{2}} Z_{+}^{*i,2}
    Z_N^{j,4}) P_L \\
&& + (Z_{-}^{i,1} Z_N^{*j,2}+\frac{1}{\sqrt{2}} (
    Z_{-}^{i,2} Z_N^{*j,3} + Z_{-}^{i,2+I} Z_N^{*j,4+I})) P_R] \\
Z_{\mu}^0-\kappa_j^{+}-\bar{\kappa}_i^{+}&:&
i\sqrt{g^2+g'^2}\gamma^{\mu} [(-\cos^2\theta_W \delta_{i j} + \frac{1}{2}
    Z_{+}^{*i,2} Z_{+}^{j,2} + Z_{+}^{*i,2+I} Z_{+}^{j,2+I}) P_L \\
&& +
    (-\cos^2\theta_W \delta_{i j} + \frac{1}{2} (Z_{-}^{i,2} Z_{-}^{*j,2}+
    Z_{-}^{i,2+I} Z_{-}^{*j,2+I} )) P_R] \\
A_{\mu}-\kappa_j^{+}-\bar{\kappa}_i^{+}&:&
-ig\sin\theta_W \gamma^{\mu} \delta_{i j} \\
H_i^0-\bar{H}_j^{+}-H_k^{+}&:&
-i A_{ecc}^{ijk} \\
H_{5+i}^0-\bar{H}_j^{+}-H_k^{+}&:&
A_{occ}^{ijk} \\
H_i^0-D_{+}^{I,j}-\bar{D}_{+}^{I,k}&:&
i\{Z_{D_I}^{*k,1} Z_{D_I}^{j,1} [\frac{1}{12} (3g^2+g'^2) A_{ve}^i -
    v_1 d^{I 2} Z_{even}^{i,1}] \\
&& + Z_{D_I}^{*k,2} Z_{D_I}^{j,2} (\frac{1}{6}
    g'^2 A_{ve}^i - v_1 d^{I 2} Z_{even}^{i,1}) \\
&& + \frac{1}{\sqrt{2}} \mu
    (d^I Z_{even}^{i,2} - d_s^I Z_{even}^{i,1}) (Z_{D_I}^{*k,1}
    Z_{D_I}^{j,2} + Z_{D_I}^{*k,2} Z_{D_I}^{j,1}) \} \\
H_{5+i}^0-D_{+}^{I,j}-\bar{D}_{+}^{I,k}&:&
\frac{1}{\sqrt{2}} \mu (d^I Z_{odd}^{i,2} + d_s^I Z_{odd}^{i,1})
    (Z_{D_I}^{*k,1} Z_{D_I}^{j,2} - Z_{D_I}^{*k,2} Z_{D_I}^{j,1}) \\
H_i^0-U_{-}^{I,j}-\bar{U}_{-}^{I,k}&:&
-i\{Z_{U_I}^{*k,1} Z_{U_I}^{j,1} [\frac{1}{12} (3g^2-g'^2) A_{ve}^i +
    v_2 u^{I 2} Z_{even}^{i,2}] \\
&& + Z_{U_I}^{*k,2} Z_{U_I}^{j,2} (\frac{1}{3}
    g'^2 A_{ve}^i + v_2 u^{I 2} Z_{even}^{i,2}) \\
&& + \frac{1}{\sqrt{2}} [\mu
    (u^I Z_{even}^{i,1} - u_s^I Z_{even}^{i,2}) -
    \epsilon_J u^I Z_{even}^{i,2+J}] \cdot \\
&& (Z_{U_I}^{*k,1} Z_{U_I}^{j,2} +
    Z_{U_I}^{*k,2} Z_{U_I}^{j,1}) \} \\
H_{5+i}^0-U_{-}^{I,j}-\bar{U}_{-}^{I,k}&:&
\frac{1}{\sqrt{2}} [\mu (u^I Z_{odd}^{i,1} + u_s^I Z_{odd}^{i,2}) -
    \epsilon_J u^I Z_{odd}^{i,2+J}] (Z_{U_I}^{*k,2} Z_{U_I}^{j,1} -
    Z_{U_I}^{*k,1} Z_{U_I}^{j,2}) \\
U_{-}^{I,j}-\kappa_i^{+}-\bar{d}^J&:&
-i C^{*JI} (g Z_{+}^{i,1} Z_{U_I}^{j,1} - u^I Z_{+}^{i,2} Z_{U_I}^{j,2}) P_L -
    i C^{*JI} d^J Z_{-}^{*i,2} Z_{U_I}^{j,1} P_R \\
D_{+}^{I,j}-\kappa_i^{+}-\bar{u}^J&:&
-i C^{IJ} (g Z_{-}^{i,1} Z_{D_I}^{j,1} + d^I Z_{-}^{i,2} Z_{D_I}^{j,2}) P_L +
    i C^{IJ} u^J Z_{+}^{*i,2} Z_{D_I}^{j,1} P_R \\
H_j^0-\kappa_i^{+}-\bar{\kappa}_k^{+}&:&
-\frac{i}{\sqrt{2}} [g Z_{-}^{k,2} Z_{+}^{i,1} Z_{even}^{j,1} + g Z_{-}^{k,1}
    Z_{+}^{i,2} Z_{even}^{j,2} + l_I Z_{-}^{k,2+I} Z_{+}^{i,2+I}
    Z_{even}^{j,1} \\
&& + (g Z_{-}^{k,2+I} Z_{+}^{i,1} - l_I Z_{-}^{k,2}
    Z_{+}^{i,2+I}) Z_{even}^{j,2+I}] P_L \\
&& -
\frac{i}{\sqrt{2}} [g Z_{-}^{*i,2} Z_{+}^{*k,1} Z_{even}^{j,1} + g Z_{-}^{*i,1}
    Z_{+}^{*k,2} Z_{even}^{j,2} \\
&& + l_I Z_{-}^{*i,2+I} Z_{+}^{*k,2+I}
    Z_{even}^{j,1} \\
&& + (g Z_{-}^{*i,2+I} Z_{+}^{*k,1} - l_I Z_{-}^{*i,2}
    Z_{+}^{*k,2+I}) Z_{even}^{j,2+I}] P_R \\
H_{5+j}^0-\kappa_i^{+}-\bar{\kappa}_k^{+}&:&
\frac{1}{\sqrt{2}} [-g Z_{-}^{k,2} Z_{+}^{i,1} Z_{odd}^{j,1} - g Z_{-}^{k,1}
    Z_{+}^{i,2} Z_{odd}^{j,2} + l_I Z_{-}^{k,2+I} Z_{+}^{i,2+I}
    Z_{odd}^{j,1} \\
&& - (g Z_{-}^{k,2+I} Z_{+}^{i,1} + l_I Z_{-}^{k,2}
    Z_{+}^{i,2+I}) Z_{odd}^{j,2+I}] P_L \\
&& -
\frac{1}{\sqrt{2}} [g Z_{-}^{*i,2} Z_{+}^{*k,1} Z_{odd}^{j,1} + g Z_{-}^{*i,1}
    Z_{+}^{*k,2} Z_{odd}^{j,2} \\
&& - l_I Z_{-}^{*i,2+I} Z_{+}^{*k,2+I}
    Z_{odd}^{j,1} \\
&& + (g Z_{-}^{*i,2+I} Z_{+}^{*k,1} + l_I Z_{-}^{*i,2}
    Z_{+}^{*k,2+I}) Z_{odd}^{j,2+I}] P_R \\
H_i^{+}-\kappa_j^{0}-\bar{\kappa}_k^{+}&:&
i [\frac{1}{\sqrt{2}} (Z_{-}^{k,2} Z_c^{i,1} + Z_{-}^{k,2+I} Z_c^{i,2+I})
    (g Z_N^{j,2} + g' Z_N^{j,1}) \\
&& - g Z_{-}^{k,1} (Z_c^{i,1} Z_N^{j,3} +
    Z_c^{i,2+I} Z_N^{j,4+I}) \\
&& - l_I Z_c^{i,5+I} (Z_{-}^{k,2+I} Z_N^{j,3} -
    Z_{-}^{k,2} Z_N^{j,4+I}) ] P_L \\
&& +
i [-\frac{1}{\sqrt{2}} Z_{+}^{*k,2} Z_c^{i,2} (g Z_N^{*j,2} + g' Z_N^{*j,1}) -
    g Z_{+}^{*k,1} Z_c^{i,2} Z_N^{*j,4} \\
&& - \sqrt{2} g' Z_{+}^{*k,2+I} Z_c^{i,5+I} Z_N^{*j,1} \\
&& - l_I Z_{+}^{*k,2+I} (Z_N^{*j,3} Z_c^{i,2+I} -
    Z_N^{*j,4+I} Z_c^{i,1}) ] P_R \\
H_i^0-d^I-\bar{d}^J&:&
\frac{i}{\sqrt{2}} d^I Z_{even}^{i,1} \\
H_{5+i}^0-d^I-\bar{d}^J&:&
\frac{1}{\sqrt{2}} d^I Z_{odd}^{i,1} \gamma_5 \\
H_i^0-u^I-\bar{u}^J&:&
-\frac{i}{\sqrt{2}} u^I Z_{even}^{i,2} \\
H_{5+i}^0-u^I-\bar{u}^J&:&
-\frac{1}{\sqrt{2}} u^I Z_{odd}^{i,2} \gamma_5
\end{eqnarray*}

\vskip 20mm
%\noindent{\bf References}

\vskip 20mm
\noindent{\Large\bf Figure captions}
\vskip 5mm

\noindent
\par
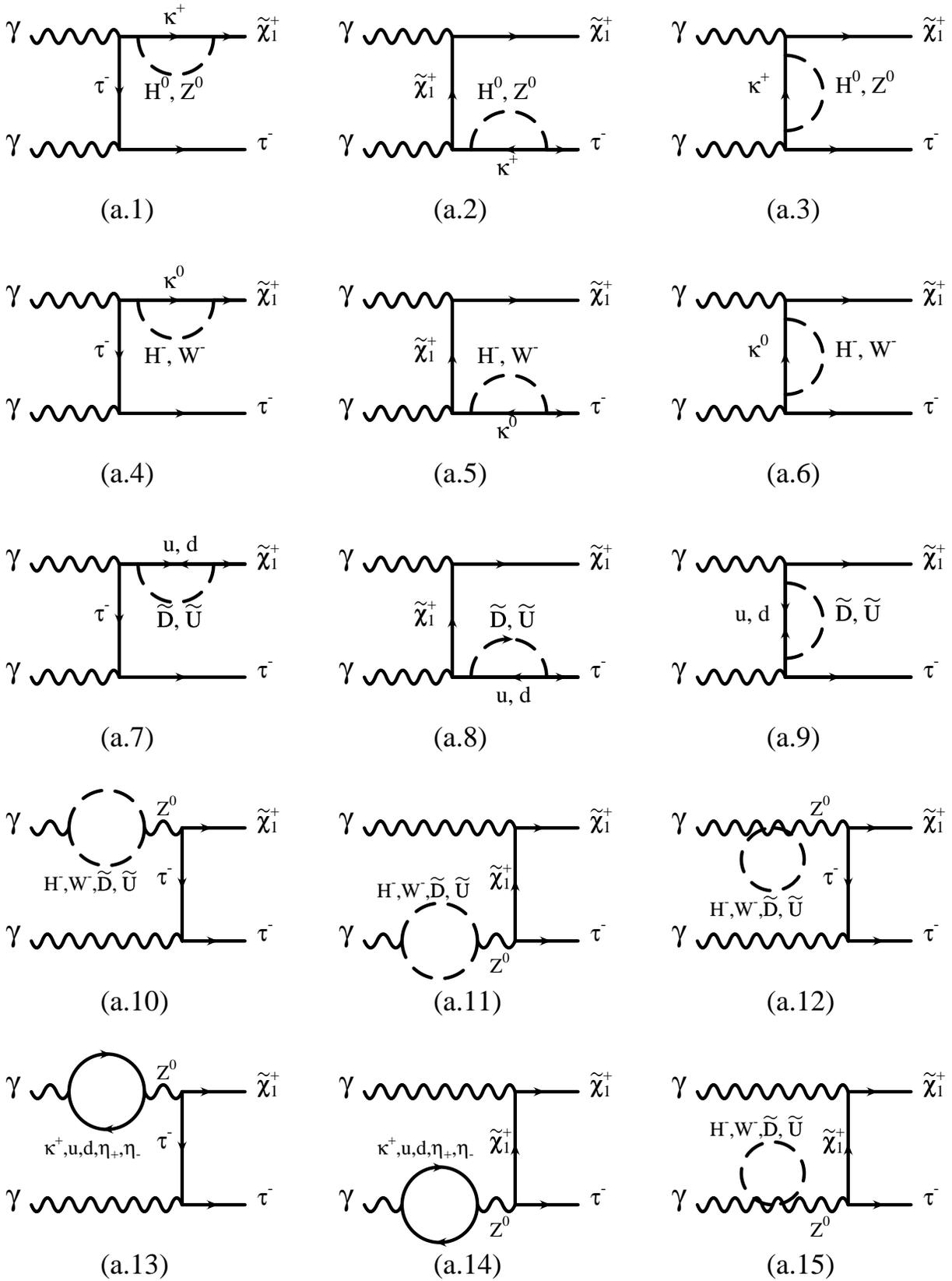
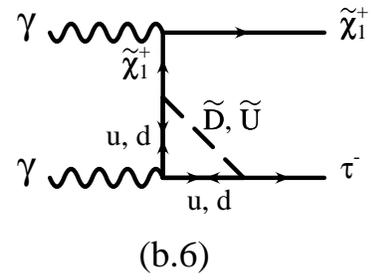
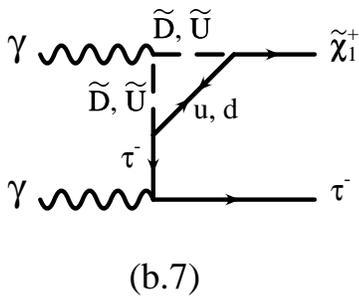
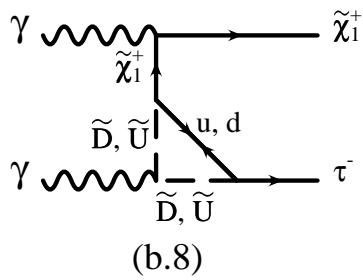
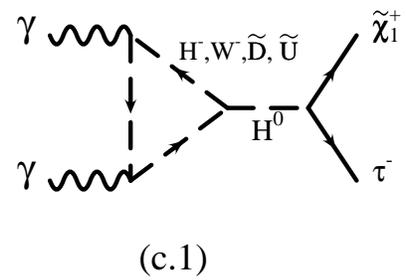
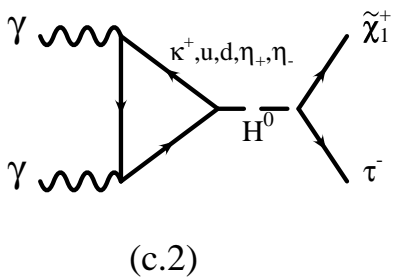
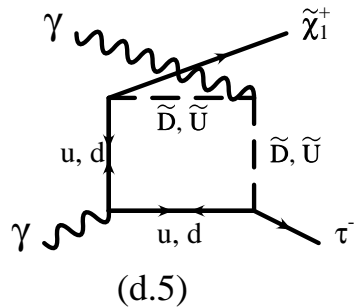
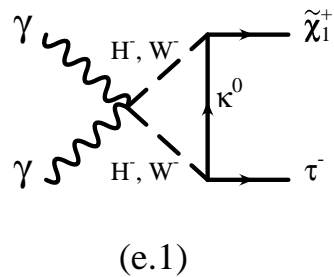
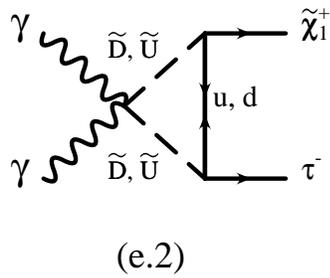
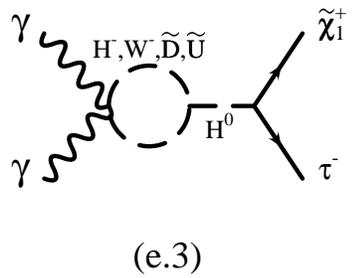
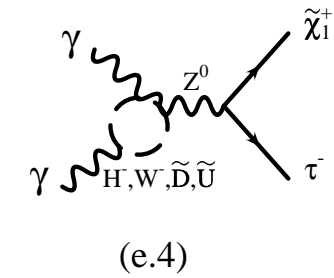
{\bf Fig.1} The ${\rlap/{R}}_p$ MSSM one-loop order diagrams
contributing to
        the $\gamma\gamma \rightarrow \tilde{\chi}_1^{+} \tau^{-}$.
        (a) self-energy diagrams; (b) vertex diagrams; (c) s-channel diagrams;
        (d) box diagrams, and (e) quartic coupling diagrams. The $H^0$,
        $H^{-}$, $\kappa^{+}$, $\kappa^0$ denote the corresponding particles
        as we have given in Eq.(12)-(19). $d$, $u$, $\tilde{D}$, $\tilde{U}$
        denote quarks and squarks respectively, where only the third
        generation is considered for convenience. The diagrams with
        incoming photons exchanged are not shown in the figures.
\par
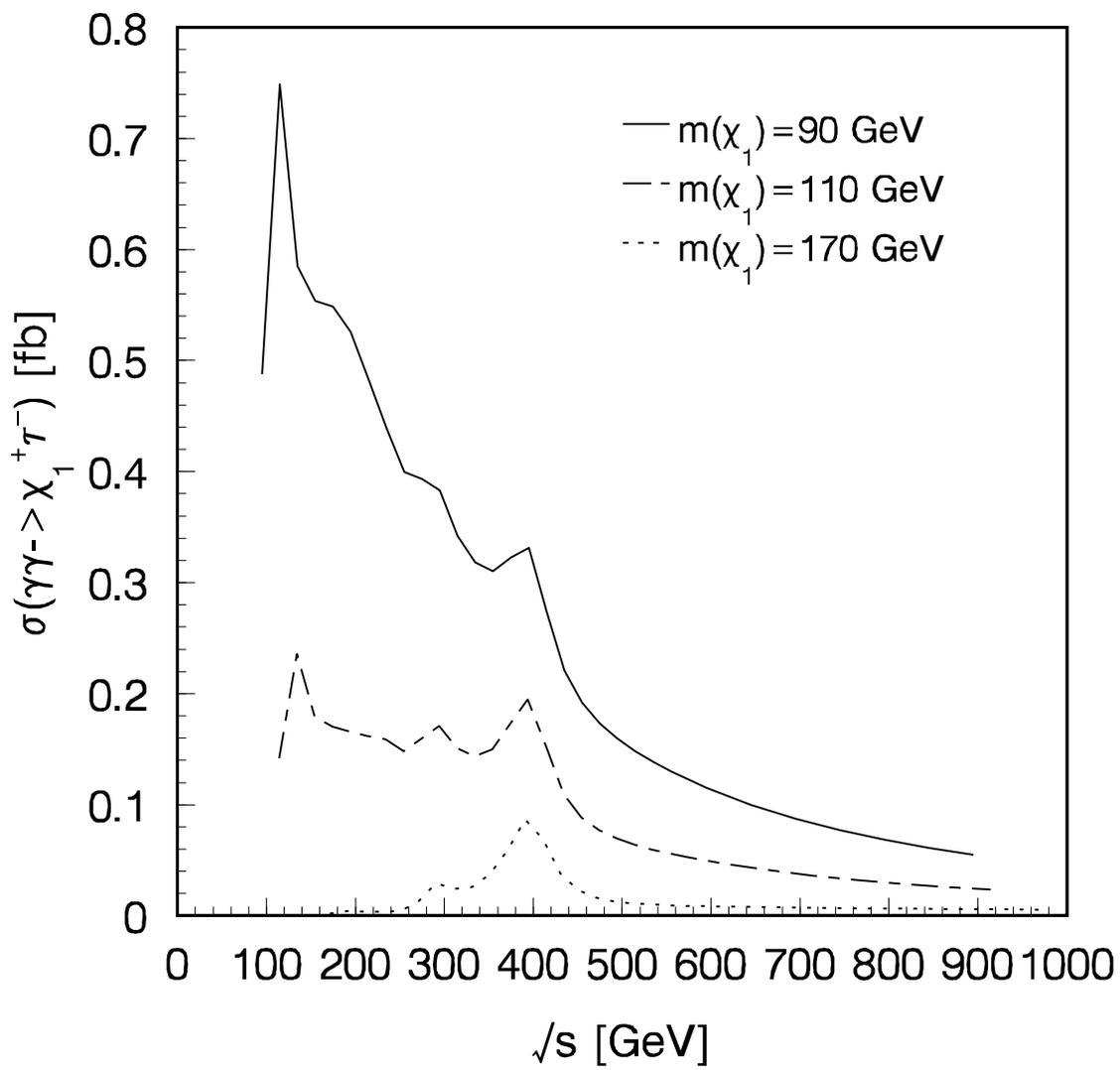
 {\bf Fig.2} The $\gamma\gamma$ subprocess cross section as functions of
        $\sqrt{\hat{s}}$, with $\mu=370~GeV$, mass of $\tilde{\chi}_1$ at
        90 GeV, 110 GeV and 170 GeV, denoted by the solid line, the dashed
        line and the dotted line respectively.
\par
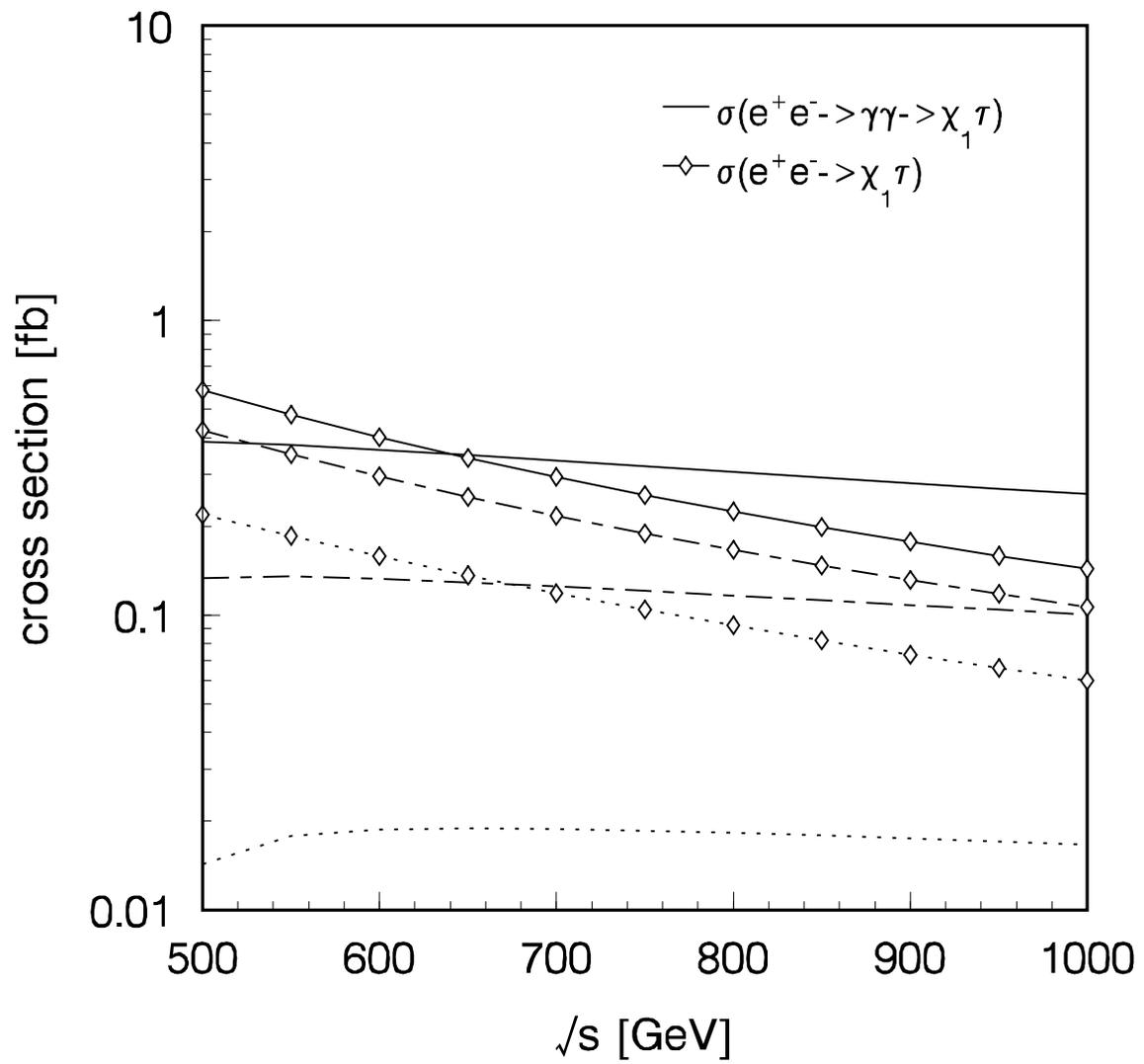
{\bf Fig.3} Both cross sections of the
$e^{+}e^{-}\rightarrow\gamma\gamma
        \tilde{\chi}_1^{+}\tau^{-}$ process and the $e^{+}e^{-}\rightarrow
        \tilde{\chi}_1^{+} \tau^{-}$ tree level process as the functions of
        $\sqrt{\hat{s}}$, with $\mu=370~GeV$. Contributions from only bilinear
        ${\rlap/ R}_p$ terms with mass of $\tilde{\chi}_1$ at
        90 GeV, 110 GeV and 170 GeV are plotted, denoted by the solid line,
        the dashed line and the dotted line respectively.
\par
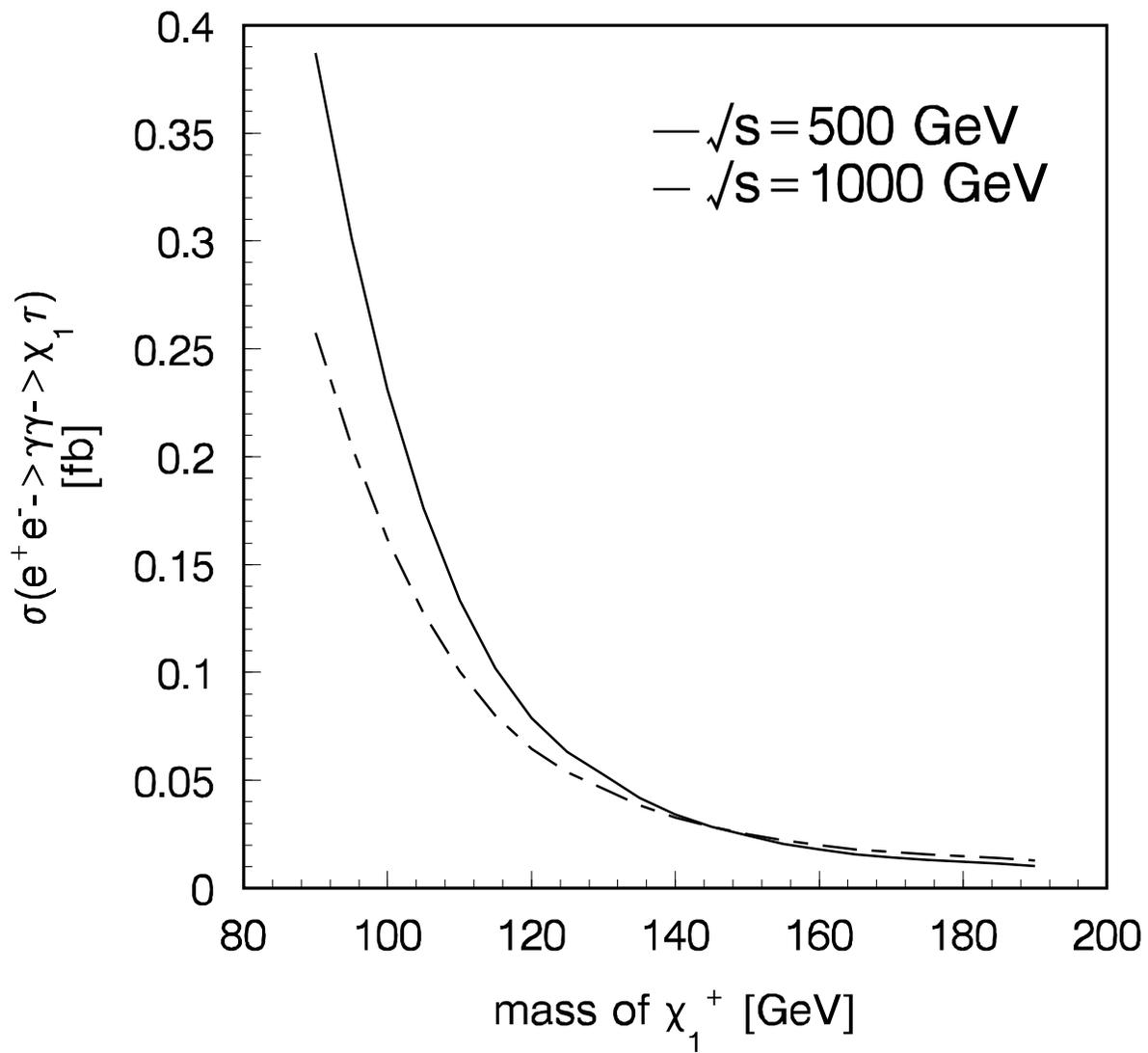
{\bf Fig.4} The cross section of the $e^{+}e^{-}$ process as
unctions of the final chargino mass $m_{\tilde{\chi}_1}$, which
ranges from 90 GeV to 200 GeV. The solid line and the dashed line
are for $\sqrt{\hat{s}}=500~GeV$ and $\sqrt{\hat{s}}=1~TeV$
respectively.
\par
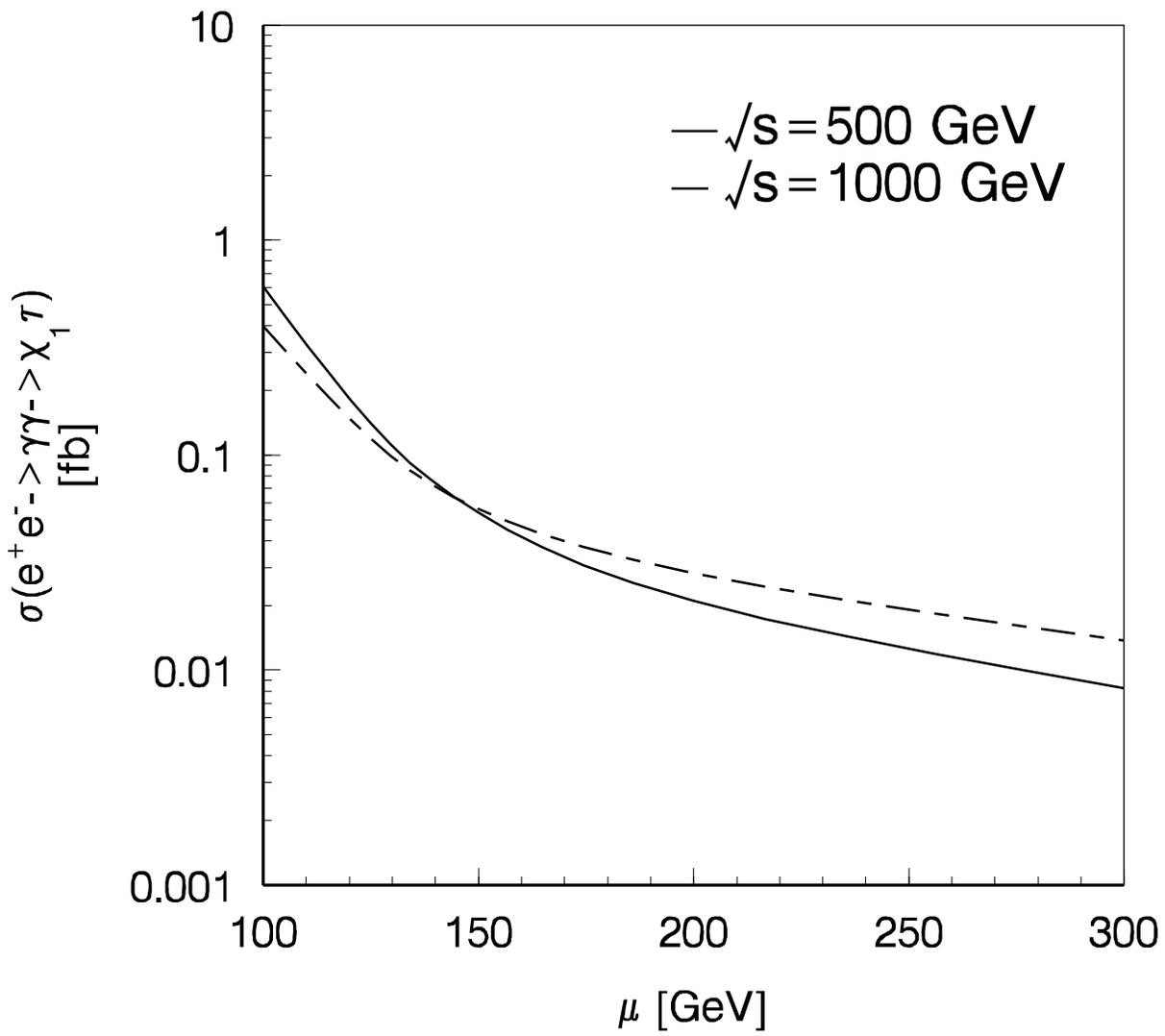
{\bf Fig.5} The cross section of the $e^{+}e^{-}$ process as functions of
        $\mu$, which is assumed positive and ranges from 100 GeV to 300 GeV.
        The solid line and the dashed line are for $\sqrt{\hat{s}} =500~GeV$
        and $\sqrt{\hat{s}}=1~TeV$ respectively.

\vskip 3mm
\noindent

\vskip 3mm
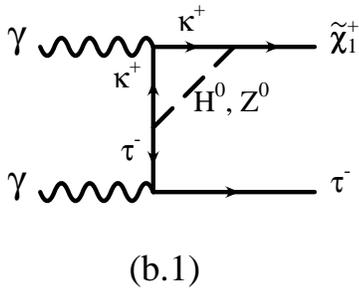
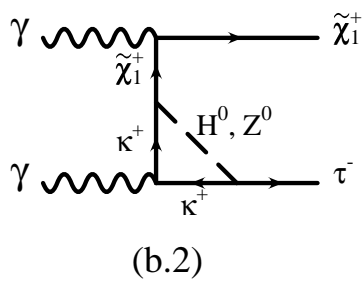
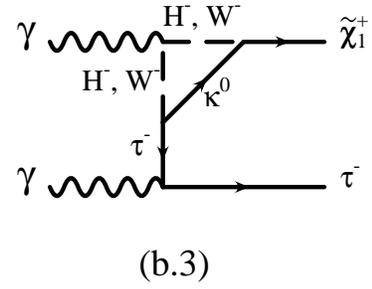
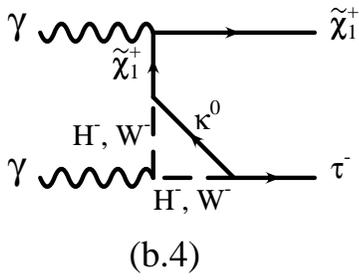
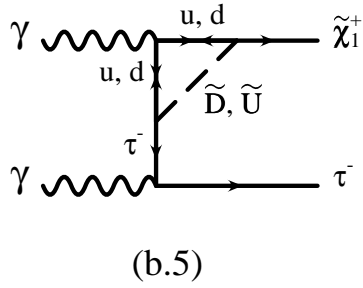
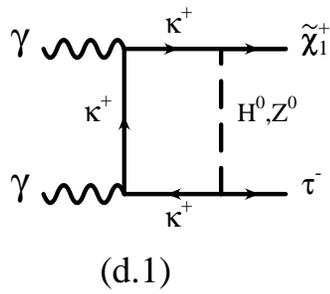
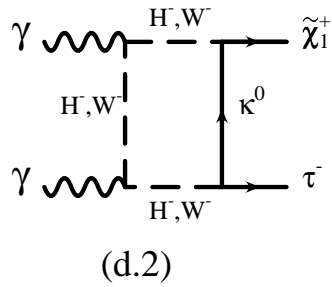
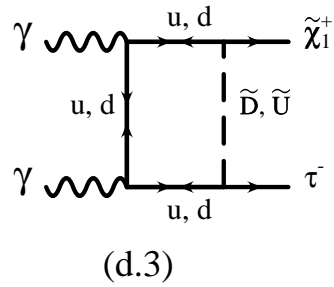
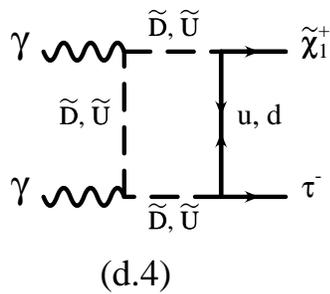
\end{document}